\documentclass[onecolumn,amsmath,amssymb,epjc3,a4paper,12pt]{svjour3}

\smartqed  % flush right qed marks, e.g. at end of proof

\usepackage[T1]{fontenc} % if needed
\RequirePackage{graphicx,bm,amsfonts,hhline,hyperref,float,eurosym,fix-cm,latexsym,epsf,mathtools,cuted,makecell,array}
\usepackage[caption = false]{subfig}
\usepackage{epstopdf,color}
\usepackage[switch]{lineno}
\usepackage{geometry}
\DeclareMathAlphabet{\mathpzc}{OT1}{pzc}{m}{it}

\begin{document}
	
	\title{Gravastar in the framework of Loop Quantum Cosmology}
	
	\author{Shounak Ghosh\thanksref{e1,addr1}
		\and
		Rikpratik Sengupta\thanksref{e2,addr2}
		\and
		Mehedi Kalam\thanksref{e3,addr2}.}
	
	\thankstext{e1}{e-mail: shounakphysics@gmail.com}
	\thankstext{e2}{e-mail: rikpratik.sengupta@gmail.com} 	
	\thankstext{e3}{e-mail: kalam@associates.iucaa.in}

	\institute{Department of Physics, Aliah University, Kolkata 700160, West Bengal, India \label{addr2}
		\and
		Directorate of Legal Metrology, Department of Consumer Affairs, Govt. of West Bengal, Malbazar, Jalpaiguri 735221, West Bengal, India \label{addr1}
	}
	
	\date{Received: date / Accepted: date}
	
	\maketitle

\begin{abstract}
In this paper we attempt to construct a regular gravastar model using the UV corrected framework of Loop Quantum Cosmology. We find that a stable gravastar model can be constructed with a number of unique features: (i) no thin shell approximation needs to be invoked to obtain solutions in the shell which can be considered to be of a finite thickness, (ii) the central singularity of a self gravitating object can be averted by a bounce mechanism, such that the interior density of the gravastar reaches a maximum critical density and cannot be raised further due to an operative repulsive force, (iii) the inherent isotropy of the effective fluid description does not prevent the formation of a stable gravastar and anisotropic pressures is not an essential requirement.  
\end{abstract}

%\pacs{04.40.Dg, 04.50.Kd, 04.20.Jb, 04.20.Dw}

\maketitle

\section{Introduction} \label{sec1}

In describing any physical phenomena or evolution of any system, the appearance of physical singularities is essentially considered to be a drawback or limitataion of the underlying physical theory. Two such physical singularities are encountered in the relativistic models of the universe\cite{Hawking} and black holes\cite{Penrose}, which are essentially curvature singularities and cannot be removed by a coordinate transformation\cite{HP}. In standard relativistic cosmology, one such curvature singularity of the Ricci type characterized by diverging Riemann curvature tensor and energy density occurs at $t=0$ marking the `beginning' of the universe and is dubbed as the initial singularity\cite{Hawking}. Similarly, relativistic black holes are charecterized by an identical curvature singularity with similar diverging parameters at $r=0$ that cannot be averted by a co-ordinate transformation\cite{Penrose}, unlike the removable singularity (not a physical one) at the Schwarzschild radius $r=2M$, where $M$ denotes the mass of the black hole\cite{Finkelstein,Kruskal,Szekeres}. In the twenty first century physics, many cosmologists and astrophysicists consider the presence of these singularities to be a limitation of the applicability of standard General Relativity (GR) to describing space-time evolution at considerably high energy densities or curvatures due to two reasons- firstly, GR is not tested at such high energy scales and secondly, it is natural to consider that quantum effects must start dominating at such high energy scales while GR is a purely classical theory.

As known since the past two and a half decades, GR is not self sufficient to explain the presently observered acceleration of the universe\cite{P1,P2,R1,R2} which is a phenomenon in the low energy domain and hence GR needs to be modified at the infra-red (IR) scales as well on large scales. The more common approach to such modifications both at the ultra-violet (UV) and IR scales has been to modify the matter sector sourcing the underlying space-time geometry. The introduction of scalar fields\cite{De1,De2,De3,De4} which find special importance in particle physics theories have been of extensive use to modify the matter sector. Besides, other exotic fluids characterized by non-linear Equation of states (EoS) have also been used widely in literature\cite{Cg1,Cg2,Cg3}. These fluids, however, often have an effective scalar field description. Also, perfect fluids with a super-negative EoS parameter\cite{Pe1,Pe2,Pe3} have found applicability in introducing IR modifications to the standard relativistic picture, with or without effective field descriptions.

Another way to modify GR that has been very popular in recent times is to modify the geometry sector characterized by the standard Einstein-Hilbert (EH) action. The EH action was modified originally in the last two decades of the previous century to introduce UV corrections but in recent times this approach has been extensively used to modify GR at both UV and IR scales and is the more preferable approach as several pathologies arising from modification of the matter source can be averted successfully in this scheme. Moreover, using this approach it might also be possible to construct a self-consistent theory of quantum gravity (QG) in the future but presently physicists are far from achieving this goal yet. However, both the early universe and black holes shall provide a natural testbed for testing the fruitfullness of any QG theory due to the extremely high energy densities and curvature that are associated with them. The two most acceptable approaches to QG theory are quite diverse in nature. Firstly, the M-theory which is a later modification leading to the unification of the five different Superstring theories via a duality mechanism reduces to the previously popular Supergravity theory at low energies and requires eleven spacetime dimensions for its quantum consistency\cite{S1,S2}. The M-theory is not fully constructed yet but it is known to be characterized by 2 and 5 dimensional objects known as 'branes' or membranes. In certain effective theories known as braneworlds, our universe is believed to be such a brane embedded in a higher dimensional bulk spacetime. Another alternative approach that does not require any additional space-time dimensions relies on the quantization of the space-time itself known as Loop Quantum Gravity\cite{L1,L2} (LQG) and has an effective description as well just like the braneworld scenario, known as Loop Quantum Cosmology\cite{LQC} (LQC). These two effective pictures namely the braneworld\cite{Randall1,Randall2} and LQC\cite{LQC} have been extremely popular in constructing cosmological as well as astrophysical models in the last two decades.

The idea of a gravastar or gravitationally vacuum condensate star was proposed by Mazur and Mottola (MM) in 2001 as a non-singular alternative to black holes as end states of stellar collapse\cite{Mazur2001}. As already discussed, the curvature singularity at the centre of the black hole was interpreted by many physicists as a drawback and hence the idea behind the proposal of gravastar was to do away with this singularity. The singularity at the event horizon in case of a static, spherically symmetric, uncharged black hole with a vacuum interior, which is the simplest case that can be considered, is characterized such that any test particle falling through it experiences nothing special due to analytic continuation through a null hypersurface where the stress-energy tensor vanishes. However, whether such continuity is permissible even on accounting quantum effects is debatable and this led MM to replace the event horizon by a thin shell\cite{Mazur2004}. The standard MM approach to construct a gravastar is as follows: The vacuum interior can be described equivalently both using dust matter where pressure vanishes or using a vacuum energy described by an EoS $p=-\rho$ which also describes Einstein's famous Cosmological Constant. So, they considered the EoS at the interior as $p=-\rho$ which would provide a repulsive effect to halt the gravitational collapse but whether it is sufficient to prevent the formation of singularity is not clear as the Null energy condition (NEC) $p+\rho\geq0$ still holds true, which essentially needs to violated to avoid singularity formation as per the famous singularity theorem of Hawking and Penrose\cite{HP}. However, as we can see the Strong energy condition (SEC) $\rho+3p\geq0$ is violated which brings repulsive gravitational effects into play. In order to account for quantum effects, following Chapline\cite{Chapline1,Chapline2}, the horizon must act as a critical surface of quantum phase transition supported by an interior violating the SEC, thereby allowing for a gravitational phase transition at the horizon leading to rearrangement of the vacuum state\cite{Gliner}. So, the interior develops into a gravitational Bose-Einstein Condensate (BEC) supported by the thin shell that has to be formed of extremely dense stiff matter characterized by an EoS $p=\rho$\cite{Zeldo1,Zeldo2} , in order to support the interior. For a detailed review on gravastar, the reader can go through\cite{RS1}.

The gravastar scenario has drawn much attention recently and stable gravastar models have been constructed in both GR as well as modified gravity frameworks\cite{G1,G2,G3,G4,G5,G6}. However, as the singularity aversion of a gravastar is a UV correction, so it would be very interesting to study gravastar in frameworks which modify GR at the UV scale. As already discussed the two main frameworks in this aspect are the braneworld and LQC, which are effective models of QG theories with and without extra dimensions, respectively. Previously, we have studied gravastar solution in the braneworld gravity with a spacelike extra dimension\cite{RS2}. A fundamental physical aspect of the braneworld models is that the standard model particles and forces are confined to the (3+1)-dimensional hypersurface known as the brane, while gravity is free to propagate in the higher dimensional bulk space-time. It is very interesting to note here that braneworld models with a time-like extra dimension\cite{VS} also exist, and these models are dual to the Randall-Sundrum models with spacelike extra dimensions. It amazingly turns out that, the modified field equations in these models are exactly identical to the ones appearing from LQC in the homogeneous and isotropic setup and can lead to similar physics from two completely different background theories. The most beautiful aspect of these theories is that they can naturally lead to singularity resolution in the cosmological context\cite{brown,LQC} and the big bang is replaced by a big bounce, where none of the concerned physical parameters diverge. This leads us to explore the posssibility of studying gravastars in the framework of LQC or braneworld models with time-like extra dimension and a vanishing bulk Weyl tensor term. We obtain the solutions for the metric potentials at both the interior and the shell and move on to analyze different physical parameters of the gravastar. We then check whether our gravastar model satisfies the different stability criteria. We finally present the physical analysis of our results in the concluding section.

\section{Mathematical formalism and solutions} \label{sec2}

The gravastar is described by a static, spherically symmetric space-time, whose line element is given by the well known form

\begin{equation}\label{eq6}
ds^2=-e^{\nu(r)}dt^2+e^{\lambda(r)}dr^2+r^2(d\theta^2+sin^2\theta d\phi^2).
\end{equation}

The modified Einstein Field Equations (EFE) for the above line element in the framework of LQC or a braneworld with time-like extra dimension in the absence of projected bulk Weyl tensor has the form

\begin{eqnarray}
&&e^{-\lambda}\left(\frac{\lambda'}{r}-\frac{1}{r^2}\right)+\frac{1}{r^2} =8 \pi \rho^{eff},\label{eq7}\\
&&e^{-\lambda}\left(\frac{\nu'}{r}+\frac{1}{r^2}\right) -\frac{1}{r^2} = 8 \pi p_r^{eff},\label{eq8}\\
&&e^{-\lambda}\left(\frac{\nu''}{2}-\frac{\lambda' \nu'}{4}+\frac{\nu'^2}{4}+\frac{\nu'-\lambda'}{2r}\right) =8 \pi p_t^{eff}.\label{eq9}
\end{eqnarray}

The additional terms arising in the geometry sector due to modification of the EH action can be expressed effectively as a modification in the matter sector, replacing the usual energy density and pressures by effective stress-energy components. For our frameworks, these effective modifications take the form

\begin{eqnarray}\nonumber
&&\rho^{eff}=\rho(r) \left( 1-\frac {\rho(r) }{\rho_c} \right) ,\\ \nonumber
&&p_r^{eff}=p_t^{eff}=p_{eff}=p(r)-{\frac {2\rho (r)  \left( p(r)+\frac{\rho(r)}{2} \right)} {\rho_c}}.\\ \nonumber
\end{eqnarray}

Here, $\rho_c$ denotes the critical density at which the universe undergoes a non-singular bounce which replaces the big bang. It is a constant model parameter. 

It is worth noting here that the effective radial and tangential pressures turn out to be equal resulting in vanishing pressure anisotropy. The more important reason for highlighting the isotropic pressure terms is as follows: We are aware that in a relativistic framework, the gravastar must have anisotropic pressures in order to be stable\cite{Cattoen}, but since here we are considering UV corrections to the relativistic EFE, it is worth exploring whether isotropic pressures can also lead to a stable and self-consistent gravastar model.

The conservation equation remains unaltered in this framework and is written as

\begin{equation} \label{eq5}
	\frac{dp}{dr}=-\frac{1}{2}\frac{d\nu}{dr}(p+\rho).
\end{equation}

We choose one of the metric potentials to be described by the Kuchowicz metric function which is a well behave potential, remaining regular for all finite radial distances and is given as \cite{Kuchowicz}
\begin{equation}
	e^{\nu(r)}=e^{Br^2+2\ln K},
\end{equation}
where $B$ and $K$ are constant parameters with the former having inverse length squared dimension and the later being dimensionless. Such a temporal metric potential has also been applied to obtain a traversable wormhole solution\cite{RS3}.

\subsection{Interior solution}

Plugging in the expression for the metric potential (6) in the modified EFE and considering the EoS for the interior $p=-\rho$, we immediately obtain the solution for the unknown radial metric potential as

\begin{equation} \label{eq11}
{{\rm e}^{-\lambda(r) }}=\frac {1}{3r} \left( -8\pi \rho_{{0}} \left( {\frac {\rho_c-\rho_0}{\rho_c}} \right) {r}^{3}+3 C_1+3r \right)
\end{equation}
where $C_1$ is an integration constant which vanishes so that the metric potential retains its regularity for all finite $r$. Invoking the conservation equation, one can easily tell that the energy density in the shell will be a constant which we have assumed to be $\rho_0$. 

The simplified metric potential can then be written as
\begin{equation} \label{eq10}
{{\rm e}^{-\lambda(r) }}=1 -\frac {8\pi \rho_{{0}}}{3} \left( {\frac {\rho_c-\rho_0}{\rho_c}} \right) {r}^{2}+3
\end{equation}

The variation of the metric potential along the radial distance has been shown in Figure 1. As we can see from the plot, the metric potential is regular and continuous throughout the interior, having a non-zero minimum at the centre of the gravastar and increases monotonically before reaching a maximum at the beginning of the shell.

\begin{figure}[!htp]
\centering
\includegraphics[width=5cm]{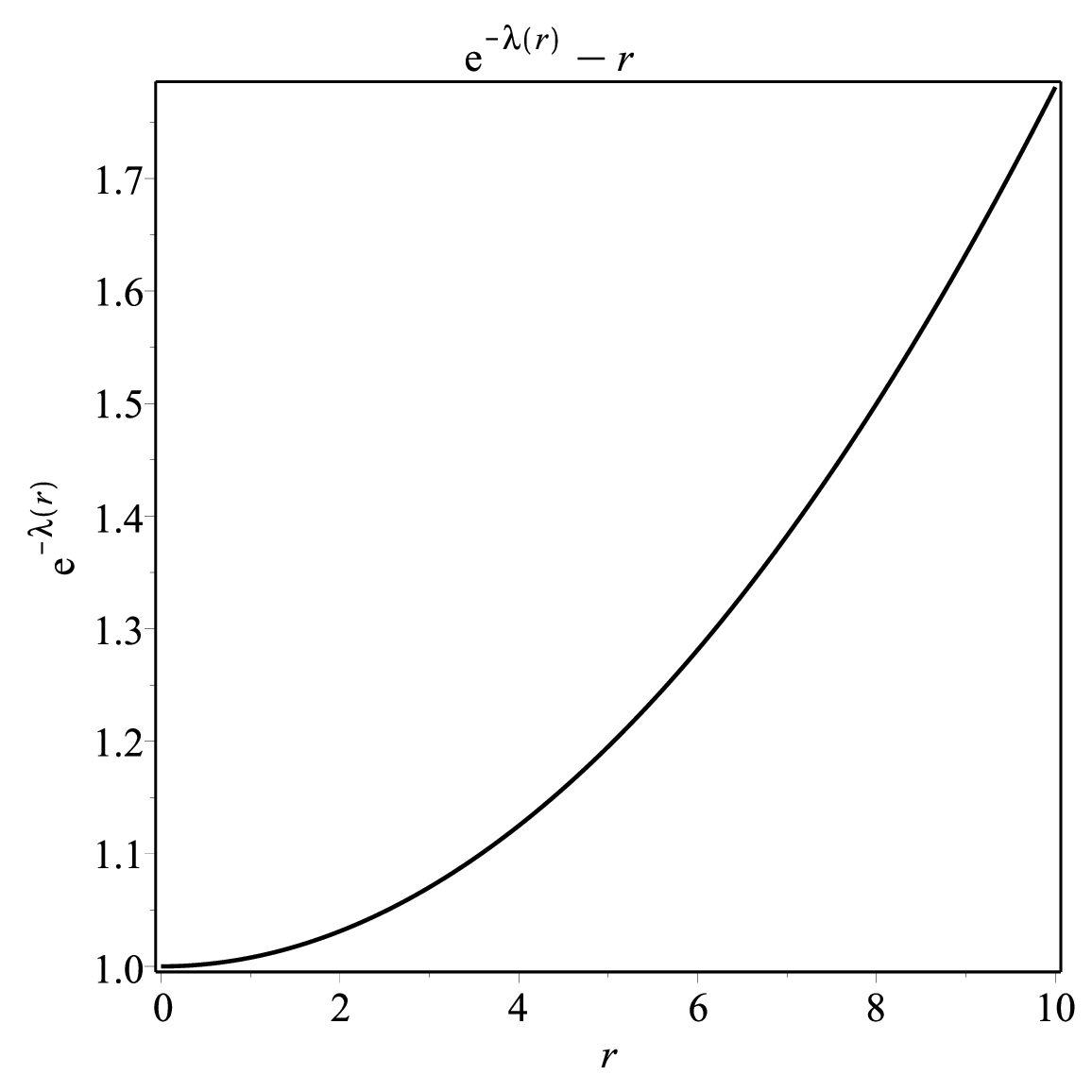}
\caption{Variation of the metric potential $g_{rr}$  w.r.t. the radial coordinate $r$.}\label{pres.}
\end{figure}

\subsection{Active gravitational mass $M(R)$}

The active gravitational mass of the interior can be computed to have the form
	\begin{equation} \label{eq15}
	M_{int}(R_1)= 4\pi\int_{0}^{R_1}\rho^{eff}r^2dr=\frac{4\pi R_1^3 \rho_0}{3} \left( 1-\frac {\rho_0 }{\rho_c} \right)
	\end{equation}

As we can see from the expression, the active gravitational mass is a constant and depends on the interior and the critical densities, the constant model parameters whose values can be estimated from the matching conditions. Here $R_1$ denotes the inner radius of the gravastar excluding the thin shell. 

\subsection{Intermediate thin Shell}

The shell is composed of stiff matter which is described by the EoS $p=\rho$. The reason for this has already been discussed in the introductory section. In case of maximum gravastar models, to solve for the unknown metric potential analytically, the thin shell approximation has to be invoked which is expressed as $e^{-\lambda}\ll 1$\cite{Mazur2001,Mazur2004}. On using this approximation the field equations are simplified to a considerable extent. However, we see that in the framework of LQC, we can solve the modified field equations in the shell analytically even without using the thin shell approximation and hence the infinitesimal thin shell can be replaced by a more realistic and physically acceptable thin shell of finite thickness. Assuming the temporal metric potential to be described by the  Kuchowicz metric function, the modified EFEs take the form given by

\begin{equation} \label{eq16}
\frac{1}{r^2}+ \left( {\frac {\lambda' }{r}}-\frac{1}{r^2} \right) {{\rm e}^{-\lambda (r) }}
=8\pi  \rho(r) \left( 1-\frac {\rho(r) }{\rho_c} \right),
\end{equation}

\begin{equation} \label{eq17}
-\frac{1}{r^2}+ \left( 2 B+\frac{1}{r^2} \right) {{\rm e}^{-\lambda (r) }}=8\pi  \left( p+{\frac {\rho \left( p+\frac{\rho}{2} \right) }{\frac{\rho_c}{2}}} \right),
\end{equation}

\begin{equation} \label{eq18}
\frac{{{\rm e}^{-\lambda (r) }}}{4} \left( 8 B+4 {B}^{2}r^2-2 Br\lambda' -{\frac{2\lambda' }{r}} \right) =8\pi  p(r)-{\frac {2\rho (r)  \left( p(r)+\frac{\rho(r)}{2} \right)} {\rho_c}}.
\end{equation}

Now, on using the EoS for the shell along with the assumed form of the metric potential $\nu(r)$ to solve the conservation equation, the unknown metric potential can be solved for analytically from the modified field equations and turns out to have the form
\begin{equation} \label{eq19}
{{\rm e}^{\lambda \left( r \right) }}={\frac {{{\rm e}^{B{r}^{2}}}{r}^{2}\rho_c{B}^{2}}{-16\pi {p_{{0}}}^{2} \left( B{r}^{2}+1 \right) {{\rm e}^{-B{r}^{2}}}+{C_2}\rho_c{B}^{2}}},
\end{equation}

where $p_0$ and $C_2$ are integration constants whose values can be estimated from the matching conditions that will be obtained using the above computed forms of the metric potentials for the interior and the shell.  

\section{Physical property}

We now analyze the various physical properties of the gravastar shell which is a junction sandwiched between the interior and the exterior spacetimes. All the physical properties of the finite thin shell must be well behaved and continuous within the shell in order to form a consistent gravastar model.

\subsection{Matter density and pressure of the shell}

Using the matter conservation equation (5) in the shell, the energy density (pressure) of the shell matter is computed to have the form
\begin{equation}\label{eq19}
p=\rho=p_0 e^{-\nu(r)}.
\end{equation}

Once we obtain the energy density for the shell matter, the effective energy density for the effective matter description involving the UV correction turns out to be
\begin{equation}\label{eq19}
\rho_{eff}=8\pi p_{{0}}{{\rm e}^{-B{r}^{2}}} \left( 1-{\frac {p_{{0}}{{\rm e}^{-B{r}^{2}}}}{\rho_c}} \right).
\end{equation}

The corresponding effective pressure can also be computed using the EoS for the shell and the energy density of the shell matter
\begin{equation}\label{eq19}
p_{eff}=8\pi  \left( p_{{0}}{{\rm e}^{-B{r}^{2}}}-{\frac{3{p_0}^{2} {{\rm e}^{-2B{r}^{2}}}^{2}}{\rho_c}} \right).
\end{equation}

Both the parameters depend on the critical density $\rho_c$ as expected. The variation of both the effective density and pressure along the radial shell thickness have been plotted in Figure 2. As we can see from the plots, both the parameters are well behaved and continuous, decreasing monotonically in a linear pattern as we move from the interior shell surface to the exterior junction.

%%%%%%%%%%%%%%%%%%%%%%%%%%%%%%%%%%%%%%%%%%%%%%%%%%%%%%%%%%%%%%%
\begin{figure}[!htp]
\centering
\includegraphics[width=5cm]{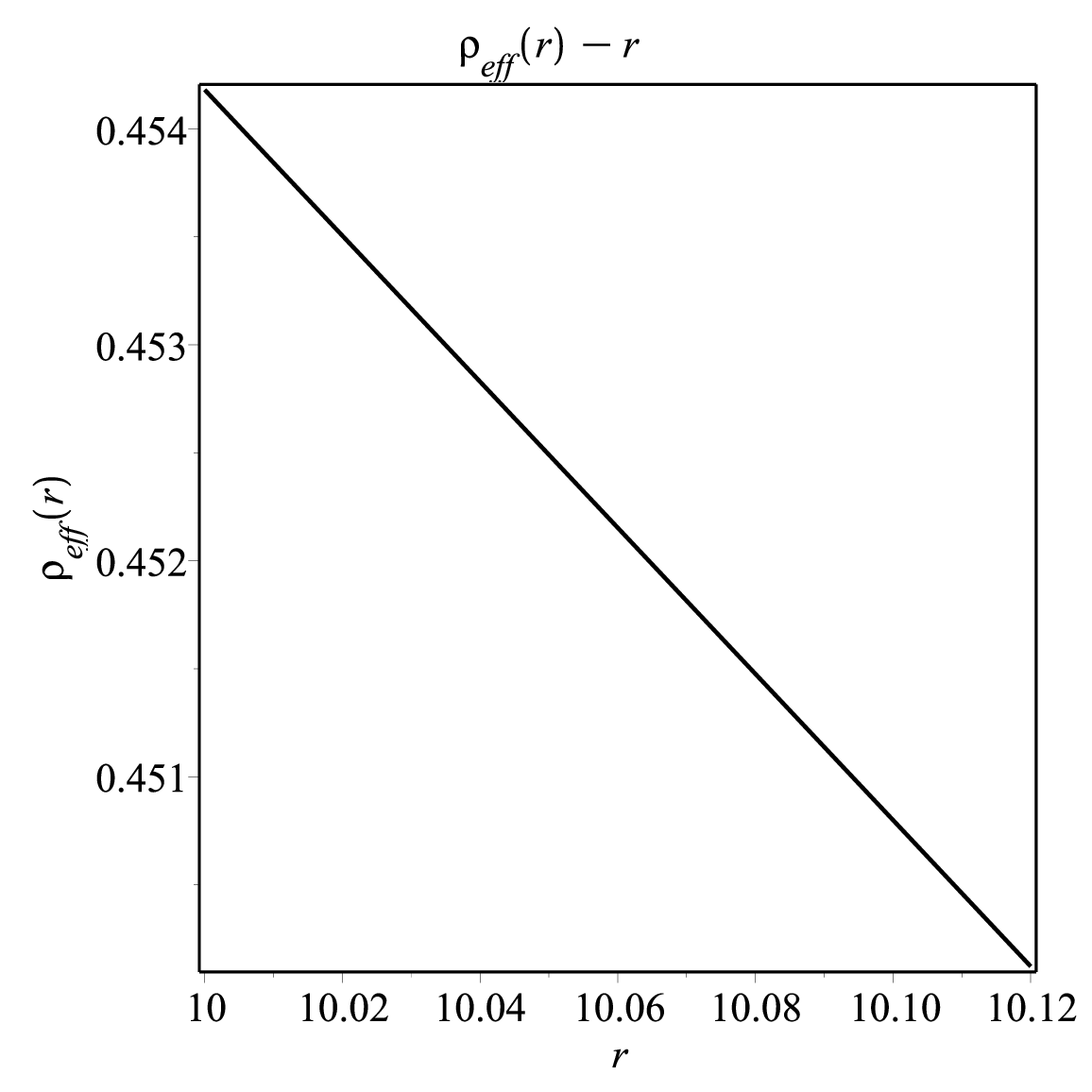}
\includegraphics[width=5cm]{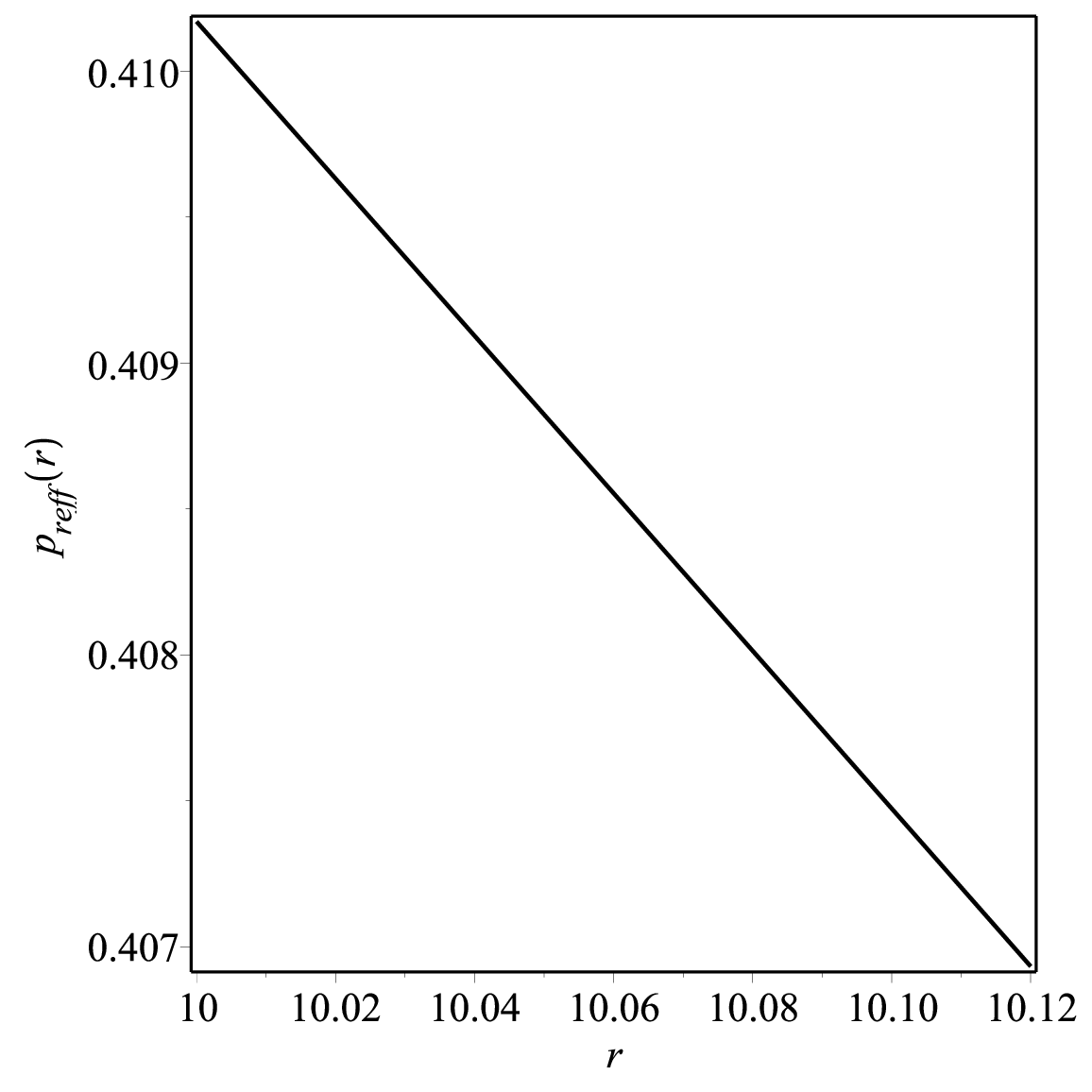}
\caption{Variation of the effective density and effective pressure  w.r.t. the radial coordinate $r$.}\label{pres.}
\end{figure}
%%%%%%%%%%%%%%%%%%%%%%%%%%%%%%%%%%%%%%%%%%%%%%%%%%%%%%%%%%%%%%%

\subsection{Energy}

In order to obtain the enrgy of the shell, we integrate the effective matter density over the surface area, yielding

\begin{eqnarray}\label{eq19}
E &=& 4\pi \int\rho^{eff}r^2 dr  \nonumber\\
&=&{\frac{8p_0{\pi }^{2} \left( 3{B}^{\frac{3}{2}}{{\rm e}^{-2B{r}^{2}}}r p_0-2{B}^{\frac{3}{2}}{{\rm e}^{-B{r}^{2}}}r\rho_c+B\sqrt {\pi } \left( -\frac{3}{4}\sqrt {2}{\rm erf} \left(\sqrt {2B}r\right)p_0+{\rm erf} \left(\sqrt {B}r\right)\rho_c \right)  \right) }{{B}^{\frac{5}{2}}\rho_c}}
\end{eqnarray}

%%%%%%%%%%%%%%%%%%%%%%%%%%%%%%%%%%%
\begin{figure*}[thbp]
\centering
\includegraphics[width=5cm]{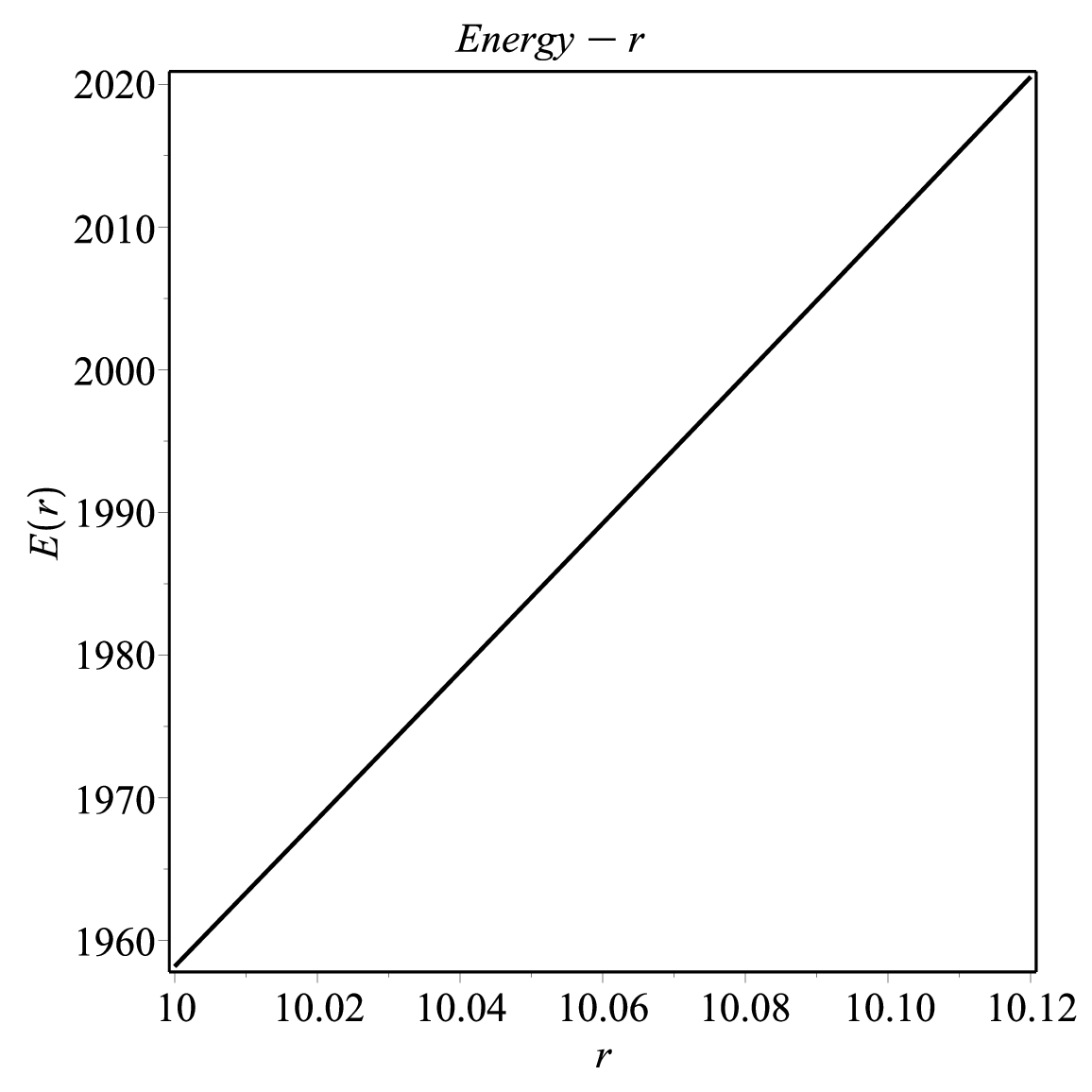}
\caption{Variation of the energy of the shell with respect to $r$.}
\end{figure*}
%%%%%%%%%%%%%%%%%%%%%%%%%%%%%%%%%%%%

The variation of the energy density of the shell along the radial expanse of the shell has been plotted in Figure 3. The energy of the shell is found to increase monotonically as we move from the inner radius to the outer radius of the gravastar.

\subsection{Entropy}

The entropy of the shell of the gravastar can be computed by using the relation
\begin{equation}\label{eq28}
S=\int_R^{R+\epsilon} 4 \pi r^2 s(r) \sqrt{e^{\lambda(r)}} dr,
\end{equation}
where the small finite quantity $\epsilon$ denotes the shell thickness and the entropy density $s(r)$ is given by the relation
\begin{equation}\label{eq29}
 s(r)=\frac{\xi^2 k_B^2 T(r)}{4\pi\hbar^2}=\frac{\xi k_B}{\hbar}\sqrt{\frac{p(r)}{2\pi}},
\end{equation}
such that $\xi$ is a dimensionless parameter. Here $k_B$ denotes the Boltzmann constant and $\hbar=\frac{h}{2\pi}$ denotes the reduced Planck's constant.  The above equation can be written down in natural units as 
\begin{equation}\label{eq30}
 s(r)=\xi\sqrt{\frac{p(r)}{2\pi}}.
\end{equation}

The entropy of the shell for our gravastar model turns out to have the form
\begin{eqnarray}\label{eq31}
S={\frac {2\sqrt {2\pi }\alpha k_{{B}}\epsilon{r}^{3}}{h}\sqrt {{\frac {p_{{0}}\rho_c{B}^{2}}{-16\pi {p_{{0}}}^{2} \left( B{r}^{2}+1 \right) {{\rm e}^{-B{r}^{2}}}+{C_2}\rho_c{B}^{2}}}}}
\end{eqnarray}
	
%%%%%%%%%%%%%%%%%%%%%%%%%%%%%%%%%%%
\begin{figure*}[thbp]
	\centering
	\includegraphics[width=5cm]{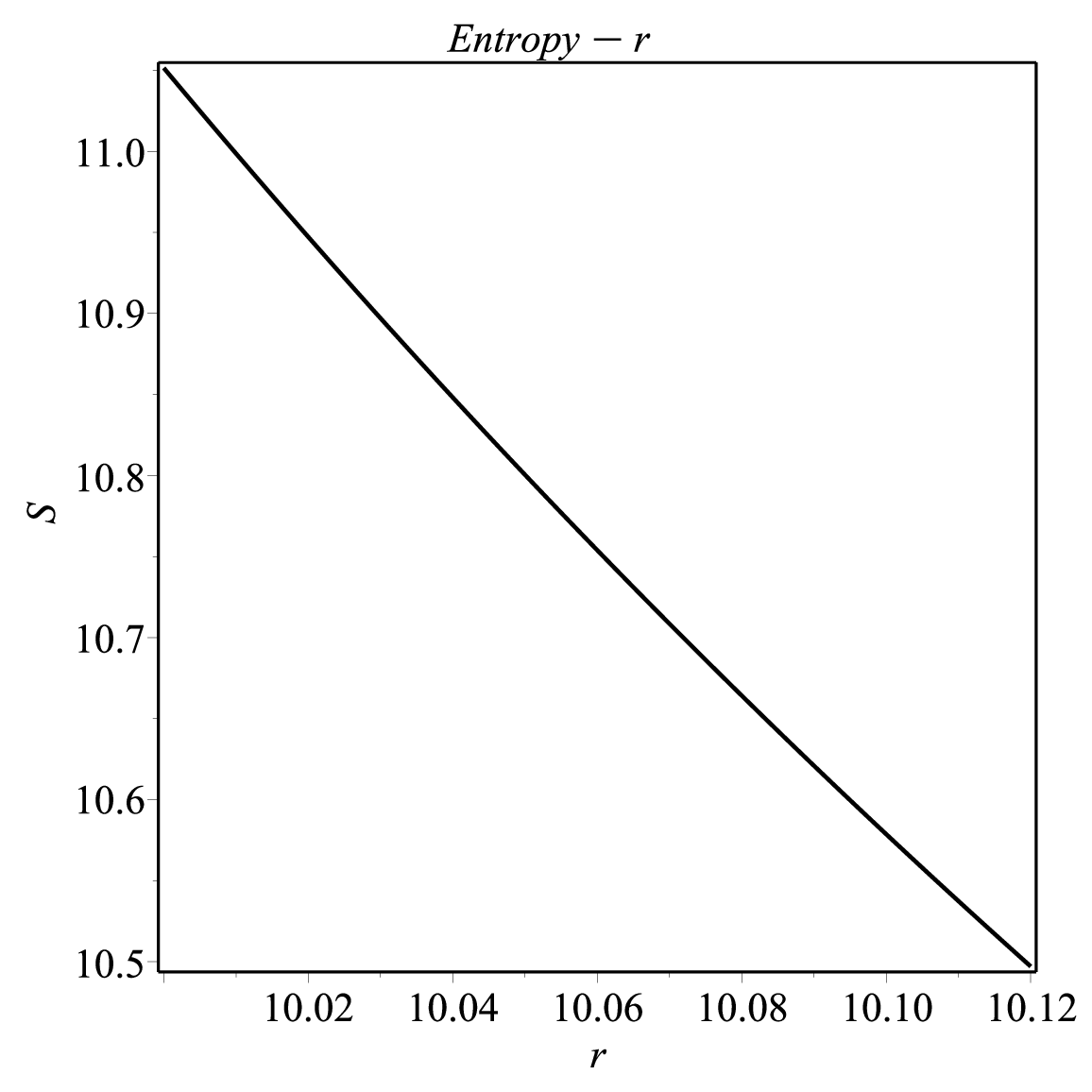}
	\caption{Variation of the entropy $(S)$ of the shell with respect to $r$.}
\end{figure*}
%%%%%%%%%%%%%%%%%%%%%%%%%%%%%%%%%%%%

The variation of the shell entropy along the radial distance across the shell has been plotted in Figure 4. We find that the the entropy is regular and well behaved and decreases monotonically as we approach the outer radius of the gravastar becoming minimum at the external surface.

\subsection{Proper thickness}

Another physical property of significant importance for a gravastar model is the proper thickness of the shell which is given by the expression
\begin{eqnarray}\label{eq32}
\ell=& &\int_{R}^{R+\epsilon}\sqrt{e^\lambda}dr =\epsilon\sqrt{e^\lambda(r)}\nonumber\\
& &=\epsilon\sqrt {{\frac {{{\rm e}^{B{r}^{2}}}{r}^{2}\rho_c{B}^{2}}{-16\pi {p_{{0}}}^{2} \left( B{r}^{2}+1 \right) {{\rm e}^{-B{r}^{2}}}+{C_2}\rho_c{B}^{2}}}}
\end{eqnarray}

%%%%%%%%%%%%%%%%%%%%%%%%%%%%%%%%%%%
\begin{figure*}[thbp]
	\centering
	\includegraphics[width=5cm]{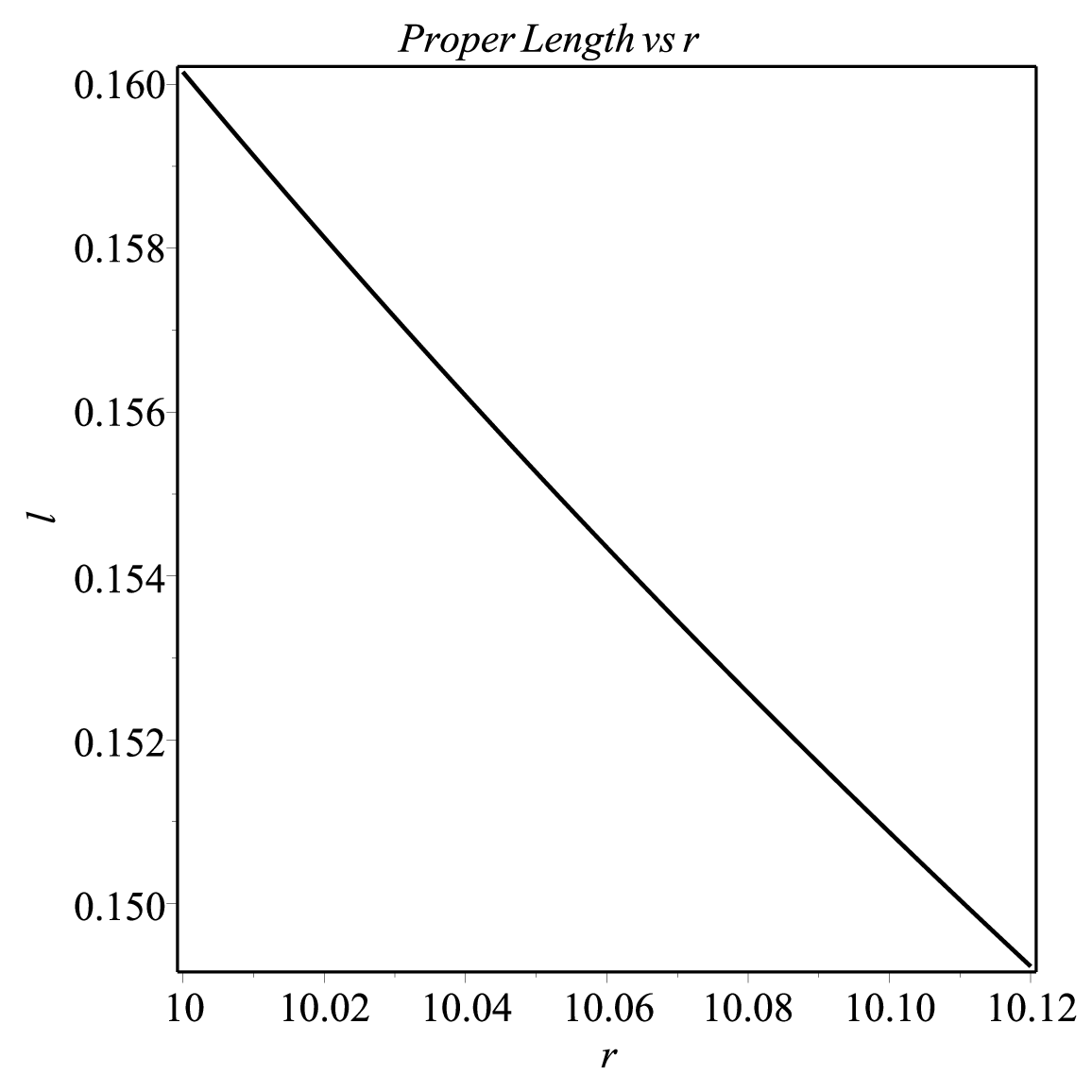}
	\caption{Variation of the proper thickness $(\ell)$ of the shell with respect to $r$.}
\end{figure*}
%%%%%%%%%%%%%%%%%%%%%%%%%%%%%%%%%%%%

The variation of the proper thickness of the shell along the radial expanse of the shell has been plotted in Figure 5. We find that, just like the entropy, the proper thickness is also maximum at the interior surface of the shell and then decreases monotonically to a non-zero minimum at the exterior surface. 

\section{Exterior spacetime}

The spacetime exterior to the gravastar is assumed to be a vacuum which can either be described by the deSitter or the Schwarzschild space-time. The simplest case in the absence of a $\Lambda$ (cosmological constant) term is described by the EoS $p=\rho=0$. Thus, using Eqs.~(2) and (3), we get
\begin{equation}\label{eq33}
    \lambda' + \nu' = 0
\end{equation}
Now, on solving the field equation for the exterior, we obtain the line-element to be given by the well known Schwarzschild metric having the form
\begin{equation}\label{eq25}
ds^2=-\left(1-\frac{2M}{r}\right)dt^2+\left(1-\frac{2M}{r}-\right)^{-1}dr^2+r^2(d\theta^2+\sin^2\theta d\phi^2),
\end{equation}
where $M$ represents the total mass of the gravastar.

\section{Boundary and matching conditions} \label{sec4}

As we know, there is presence of stiff matter on the gravastar surface to support the formation of gravitational BEC in the interior. As a result of this, an extrinsic discontinuity is present resulting in the generation of an intrinsic surface energy density and surface pressure, thus allowing the inner surface of the gravastar to act as a boundary between the shell and interior regions and the outer surface of the gravastar to act as a boundary between the shell and exterior regions. This results in the gravastar structure being a geodesically complete manifold and therefore the continuity of the metric potentials at the interior and exterior surfaces of the gravastar formed by the interior-shell and exterior-shell interfaces, respectively, yields the matching conditions. Following the Darmois-Israel mechanism, the components of the intrinsic stress-energy tensor are obtained~\cite{Darmois,Israel}.  

\begin{equation}\label{eq41}
    S_{j}^{i}= - \frac{1}{8\pi}\left[K_{j}^{i}- \delta_{j}^{i} K_{k}^{k}\right].
\end{equation}

Here, the discontinuity in the second fundamental form can be expressed as
\begin{equation}\label{eq42}
    K_{ij}= K_{ij}^{+} -  K_{ij}^{-},
\end{equation}

where the second elemental form is given by
\begin{equation}\label{eq43}
K_{ij}^{\pm} = - \eta_{\nu}^{\pm}\left[\frac{\partial^{2} X_{\nu}}{\partial \zeta^{i} \partial \zeta^{j}} + \Gamma^{\nu}_{\alpha\beta}\frac{\partial X^{\alpha}}{\partial \zeta^{i}} \frac{\partial X^{\beta}}{\partial \zeta^{j}} \right]. 
\end{equation}

$\eta_{\nu}^{\pm} = \pm \left|g^{\alpha \beta} \frac{\partial f}{\partial X^{\alpha}}
\frac{\partial f}{\partial X^{\beta}} \right|^{\frac{1}{2}} \frac{\partial f}{\partial X^{\nu}}$ represents the unit normal vector, with $\eta^{\nu}\eta_{\nu} = 1$.
$\zeta^{i}$ denotes the intrinsic shell coordinate and the parametric equation of the shell is written as $f(X^{\alpha}(\zeta^{i})) = 0$,the $+$ and $-$ signs corresponding  to the space-time at the exterior and interior of the gravastar,respectively.

Due to the spherical symmetry of the gravastar, the components of the surface energy momentum tensor has the form $S_{ij}= diag (-\Sigma,P, P, P)$, where $\Sigma$ and $P$ denote the surface energy density and surface pressure, respectively. 

The surface energy density can be computed as

\begin{eqnarray}\label{eq33}
   \Sigma & & =-\frac{1}{4\pi r}\bigg[\sqrt{e^\lambda}\bigg]_-^+ \nonumber\\
   & =&-{\frac {1}{4\pi r} \left( \sqrt {1-\frac{2M}{r}}-\sqrt{1 -\frac {8\pi \rho_{{0}}}{3} \left( {\frac {\rho_c-\rho_0}{\rho_c}} \right) {r}^{2}+3} \right) }.
	\end{eqnarray}

%%%%%%%%%%%%%%%%%%%%%%%%%%%%%%%%%%%
\begin{figure*}[thbp]
	\centering
	\includegraphics[width=5cm]{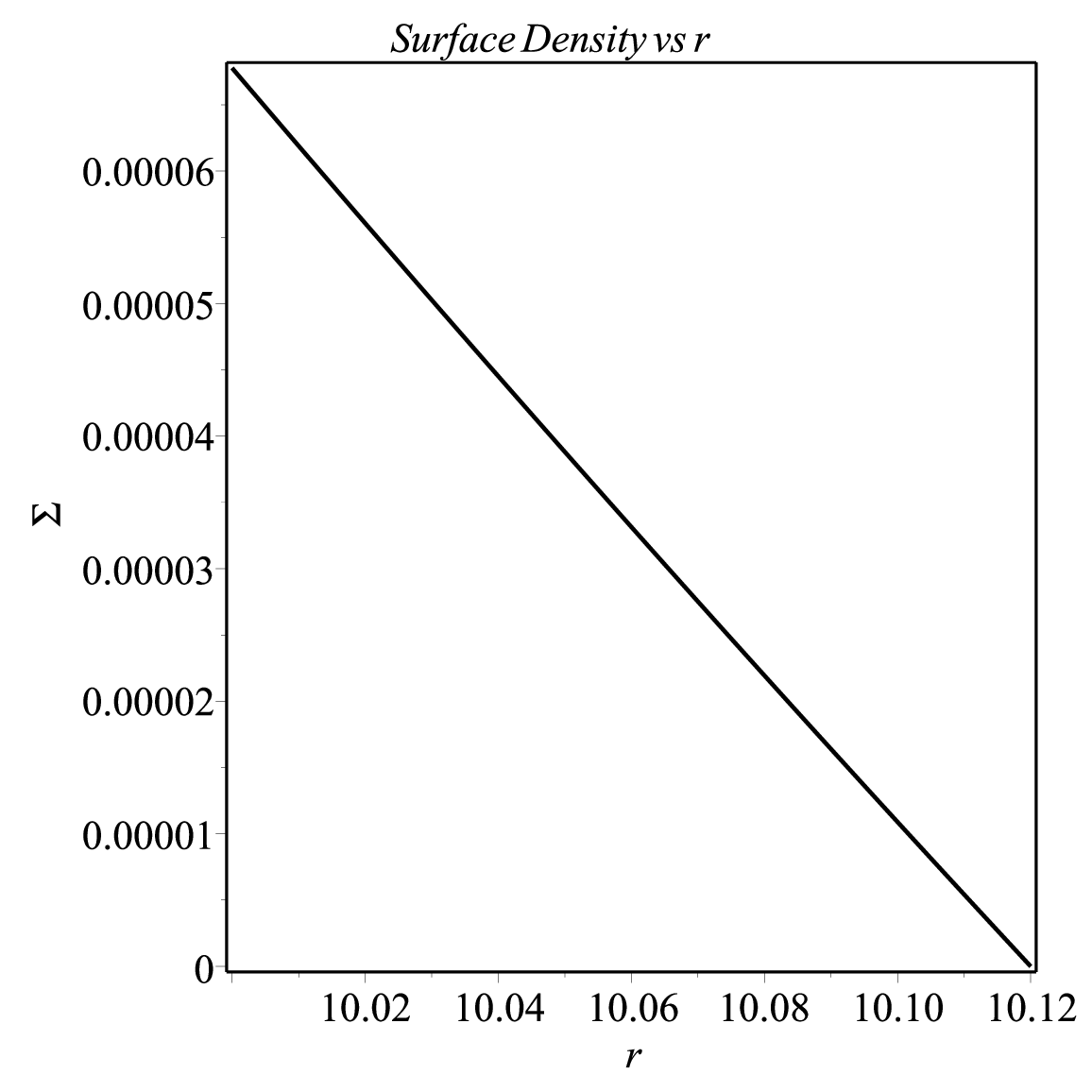}
	\caption{Variation of the Surface energy density of the shell with respect to $r$.}
\end{figure*}
%%%%%%%%%%%%%%%%%%%%%%%%%%%%%%%%%%%%

The variation of the surface energy density along the radial expanse of the thin shell has been plotted in Figure 6. The surface energy density turns out to be positive throughout the shell and is well behaved, being maximum at the interior surface of the gravastar. As we move towards the exterior surface, the surface density decreases and is minimum at the exterior surface of the gravastar.

The surface pressure of the gravastar turns out to have the form  

\begin{eqnarray}\label{eq34}
\mathcal{P} & =&\frac{1}{16\pi } \bigg[\bigg(\frac{2-\lambda^\prime r}{r}\bigg) \sqrt{e^{-\lambda}}\bigg]_-^+ \nonumber\\
&=&\frac{1}{4}{\frac {r-M}{\pi {r}^{2}}{\frac {1}{\sqrt {1-{\frac{2M}{r}}}}}}-{\frac{ \left( 8 \rho_0{r}^{2} \left( \rho_0-\rho_c \right) \pi +3\rho_c \right) }{4\sqrt{3\rho_c}\pi r {\sqrt {8\pi {r}^{2}{\rho_0}^{2}-8\pi {r}^{2}\rho_0\rho_c+3\rho_c}}}}
\end{eqnarray}

%%%%%%%%%%%%%%%%%%%%%%%%%%%%%%%%%%%
\begin{figure*}[thbp]
\centering
\includegraphics[width=5cm]{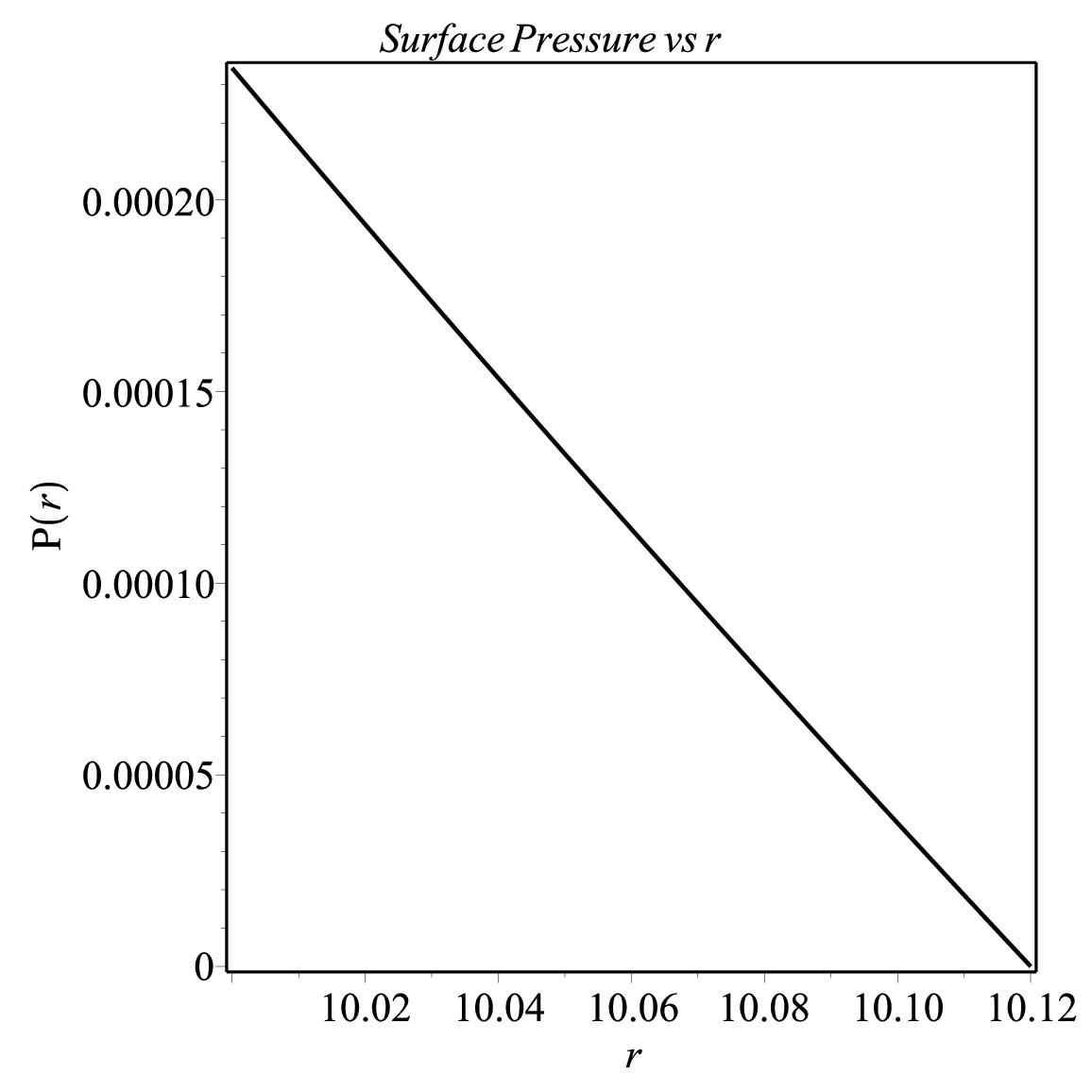}
\caption{Variation of the Surface pressure of the shell with respect to $r$.}
\end{figure*}
%%%%%%%%%%%%%%%%%%%%%%%%%%%%%%%%%%%%

The variation of the surface pressure along the radial direction across the shell is plotted in Figure 7. Just like the surface energy density, the surface pressure is also well behaved throughout the shell, having a maximum at the interior surface and decreasing radially outwards. 

The mass of the thin shell (using Eq. (\ref{eq33})) turns out to be
\begin{eqnarray}
m_s&=&4\pi r^2 \Sigma \nonumber\\
m_{{s}}&=&-r \left( \sqrt {1-2{\frac {M}{r}}}-1/3\sqrt {3}\sqrt {{\frac {1}{r} \left( -8\pi \rho_0 \left( {\frac {\rho_c-\rho_0}{\rho_c}} \right) {r}^{3}+3r \right) }} \right)
\end{eqnarray}

The total mass of the shell has the form 

\begin{equation}\label{20}
M_{Tot}=-\frac{1}{2} \left(  \left( -{\frac {m_{{s}}}{R_2}}+\frac{1}{3}\sqrt {3 \left( -8{\frac {\pi \rho_0{R_2}^{2}}{\rho_c}}+3{\rho_c}^{-1} \right) \rho_c+24{\frac {{R_2}^{2}\pi {\rho_0}^{2}}{\rho_c}}} \right) ^{2}-1 \right) R_2,
\end{equation}
where $R_2$ denotes the radius of the exterior surface of the gravastar including the shell. The finite thickness of the gravastar $\epsilon=R_2-R_1$.

\subsection{Matching Condition}

The analytic continuation of the metric function $\lambda$ at the interior boundary of the gravastar and the analytic continuation of $\nu$ at the exterior boundary of the gravastar provides us with two of the matching conditions. The third matching condition arises with the analytic continuation of the first radial derivatives of the metric function $\nu$ at the exterior boundary. The three matching conditions are accompanied by the two boundary conditions which comprise of the vanishing of the surface energy density and surface pressure at the exterior boundary of the gravastar. These five conditions shall be used to estimate the values of the five unknown model parameters, namely the mass of the gravastar $M$, the constant interior density of the gravastar $\rho_0$, the Kuchowicz parameters $B$ and $K$ and the integration constant $C_2$. Realistic values have been considered for the parameters $R_1$ and $R_2$ describing the interior and exterior radii of the gravastar, the integration constant $p_0$ and the critical density $\rho_c$. In order to calculate the arbitrary constants we have matched various physical parameters at the boundary formed between the interior-shell and shell-exterior of the gravastar at $r=R_1$ and $r=R_2$ respectively. At the boundary of the different regions, the metric functions and their derivatives must be continuous. So we matched $g_{rr}$ at $r=R_1$, $g_{tt}$ and $\frac{\partial g_{tt}}{\partial r}$ at $r=R_2$. Again at the boundary of shell and exterior the surface energy density and surface pressure must vanish.

\begin{eqnarray}
	(i) &&g_{tt}^{shell}|_{r=R_2}=g_{tt}^{ext}|_{r=R_2}\nonumber\\
	&\Rightarrow&1-{\frac{2m}{R_2}}={{\rm e}^{B{R_2}^2}}{K}^2 
\end{eqnarray}

\begin{eqnarray}
	(ii) &&\bigg(\frac{\partial{g_{tt}}}{\partial r}\bigg)^{shell}|_{r=R_2}=\bigg(\frac{\partial{g_{tt}}}{\partial r}\bigg)^{ext}|_{r=R_2}\nonumber\\
	&\Rightarrow&2BR_2={\frac{2m}{{R_2}^2}}
\end{eqnarray}

\begin{eqnarray}
	(iii)	&&g_{rr}^{int}|_{r=R_1}=g_{rr}^{shell}|_{r=R_1}\nonumber\\
		&\Rightarrow&{\frac{{{\rm e}^{B{R_1}^{2}}}{R_1}^2\rho_c{B}^2}{-16\pi {p_{{0}}}^2 \left( B{R_1}^2+1 \right) {{\rm e}^{-B{R_1}^2}}+{C_2}\rho_c{B}^2}}=\left( -{\frac{8\pi \rho_0{R_1}^2}{3\rho_c}}+3{\rho_c}^{-1} \right) \rho_c+{\frac{8{R_1}^2\pi {\rho_0}^2}{3\rho_c}}
\end{eqnarray}

\textit{Boundary Conditions}

\begin{eqnarray}
	(iv) &&\Sigma=0|_{r=R_2}=0\nonumber\\
	     &\Rightarrow&-{\frac{1}{4\pi R_2} \left( \sqrt {1-{\frac{2m}{R_2}}}-\frac{1}{9}\sqrt{3 \left( -{\frac{8\pi \rho_0{R_2}^2}{\rho_c}}+3{\rho_c}^{-1} \right) \rho_c+{\frac {24{R_2}^2\pi {\rho_0}^2}{\rho_c}}} \right) }=0
\end{eqnarray}

\begin{eqnarray}
	(v) &&\mathcal{P}=0|_{r=R_2}=0\nonumber\\
	&\Rightarrow&\frac{1}{4}{\frac {R_2-M}{\pi {R_2}^{2}}{\frac {1}{\sqrt {1-{\frac{2M}{R_2}}}}}}-{\frac{ \left( 8 \rho_0{R_2}^{2} \left( \rho_0-\rho_c \right) \pi +3\rho_c \right) }{4\sqrt{3\rho_c}\pi R_2 {\sqrt {8\pi {R_2}^{2}{\rho_0}^{2}-8\pi {R_2}^{2}\rho_0\rho_c+3\rho_c}}}}=0
\end{eqnarray}

Using the above conditions along with physical realistic values of the model parameters as $R_1 = 10 \ km,\  \rho_c = 0.41 m^4, R_2 = 10.12 \ km, p_0 = 0.028000027 m^4$ , we can obtain the unknown constant parameters as $M = 4.048000027M_\odot,\  \rho_0 = 0.4109303066 m^4 ,\ B = 0.003905700786 km^{-2},\  K = 0.3661475185,  C_2 = 6011.888170$ by using the matching and boundary conditions given by Equations (32)-(36). These obtained values of the different parameters have been used to draw the plots. The energy densities and pressures are in units of Planck mass. It is to be noted here that since we have considered a finite thin shell so a shell thickness of $0.12km$ is a realistic one for our model. There is a special significance of the fact that the interior density of the gravastar turns out to be equal to the critical density, which we shall discuss elaborately in the concluding section.

\section{Stability}

We perform a three-fold stability check to confirm the stability of our gravastar model in the framework of LQC. We first provide a cursory check by obtaining the variation of the surface redshift within the gravastar shell. The shell of the gravastar is composed of stiff matter to support the interior. Such matter is in the extreme limit of causality as the sound speed is extremely high in such matter medium and hence it is essential to perform multiple stability checks for our gravastar model to be considered physically viable. Following the surface redshift we move on to check the validity of the different energy conditions within the shell. This is a very essential stability check. Finally we check the Herrera's cracking condition or the variation of sound speed squared (due to inherent absence of pressure anisotropy in our model) in the effective matter description within the shell.

\subsection{Surface Redshift}

The surface redshift for our gravastar model can be computed as 

\begin{eqnarray}\label{eq35}
	Z_{s}&=&-1+\frac{1}{\sqrt {g_{\it tt}}} =-1+{\frac {1}{\sqrt {{K}^{2}{{\rm e}^{Br^2}}}}}.
\end{eqnarray}

The variation of the surface redshift along the radial distance $r$ from the ineer radius of the gravastar to its outer radius has been plotted in Figure 8. For any self-gravitating compact object, the surface redshift $Z_s\leq2$~\cite{Buchdahl1959,Straumann1984,Bohmer2006} in the absence of a $\Lambda$ term. As already discussed, we donot consider any $\Lambda$ term in the framework of LQC for our analysis. As we find from the plot using the constant model parameter values obtained using the matching and boundary conditions, the surface redshift is maximum at the interior boundary of the gravastar and has a non-zero minimum at the exterior boundary, decreasing monotonically across the radial expanse of the shell. At all points along the shell, it is well behaved and the value is well less than the desired upper limit, thus assuring the stability of our gravastar model.  

%%%%%%%%%%%%%%%%%%%%%%%%%%%%%%%%%%%
\begin{figure*}[thbp]
	\centering
	\includegraphics[width=5cm]{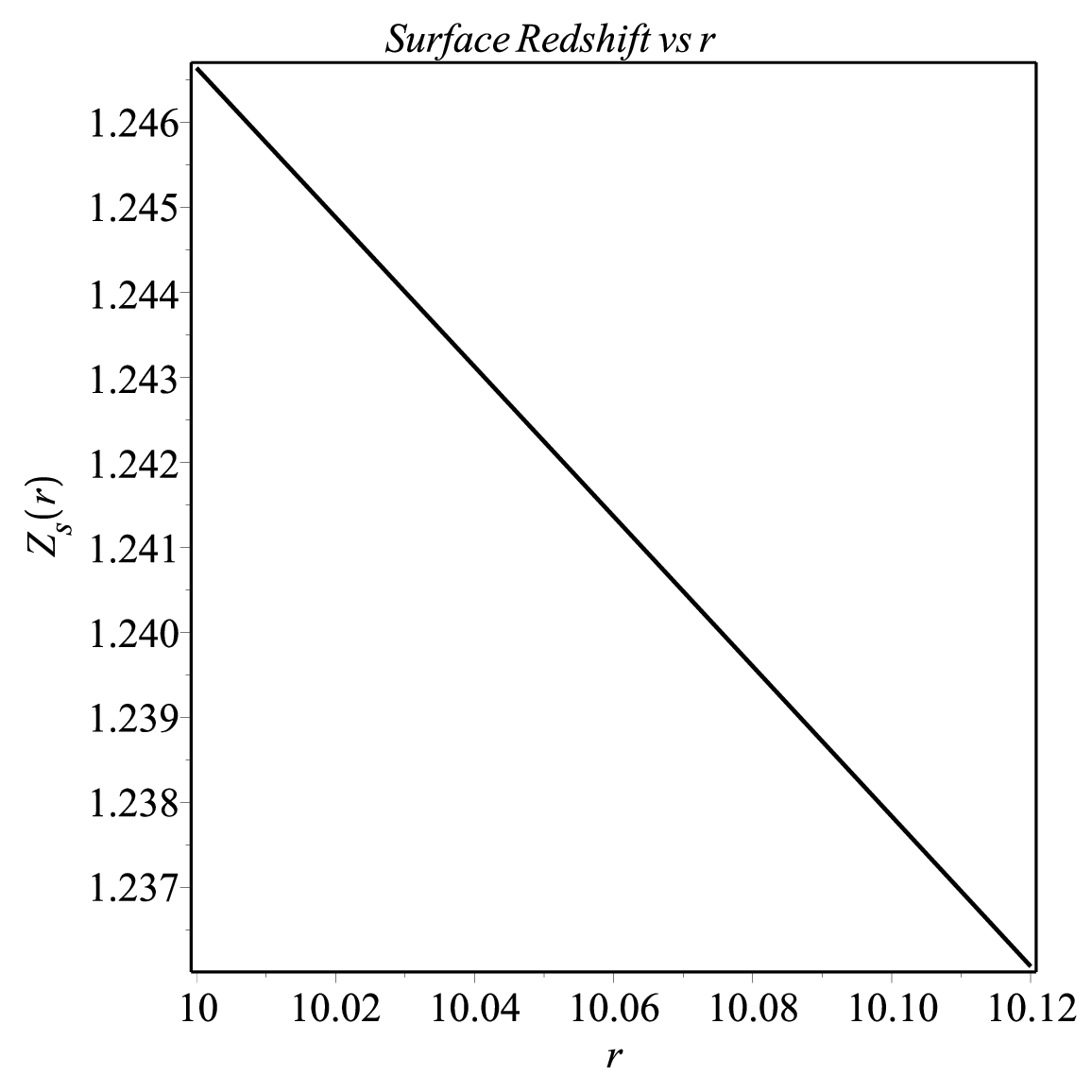}
	\caption{Variation of the Surface redshift of the shell with respect to $r$.}
\end{figure*}
%%%%%%%%%%%%%%%%%%%%%%%%%%%%%%%%%%%%

\subsection{Energy conditions}

As discussed before, in LQC, the modification in the geometry sector of the field equations can effectively be expressed as a modification in the matter sector leading to an ``effective matter" description. So, the energy conditions must be satisfied by this effective matter rather than the usual matter source. The energy conditions in terms of the effective stress-energy tensor are as follows  

\begin{eqnarray}
&&NEC: \rho^{eff} \geq 0,   \nonumber\\      
&&WEC: \rho^{eff}+p^{eff} \geq 0 , \nonumber\\ 
&&SEC: \rho^{eff}+3 p^{eff}\geq 0,   \nonumber\\     
&&DEC: \rho^{eff}-|p^{eff}| \geq 0  \label{eq39}
\end{eqnarray}

The variation of the different linear combinations of the effective stress-energy components corresponding to the different energy conditions along the radial distance $r$ across the finite thin shell have been plotted in Figure 9. We see that all the physically relevant combinations involving the effective quantities have non-zero values for the entire expanse of the shell. Thus, all the energy conditions are satisfied by the gravastar in the effective matter description in both the equaivalent frameworks.  

%%%%%%%%%%%%%%%%%%%%%%%%%%%%%%%%%%%%%%%%%%%%%%%%%
\begin{figure}[!htp]
\centering
\includegraphics[width=5cm]{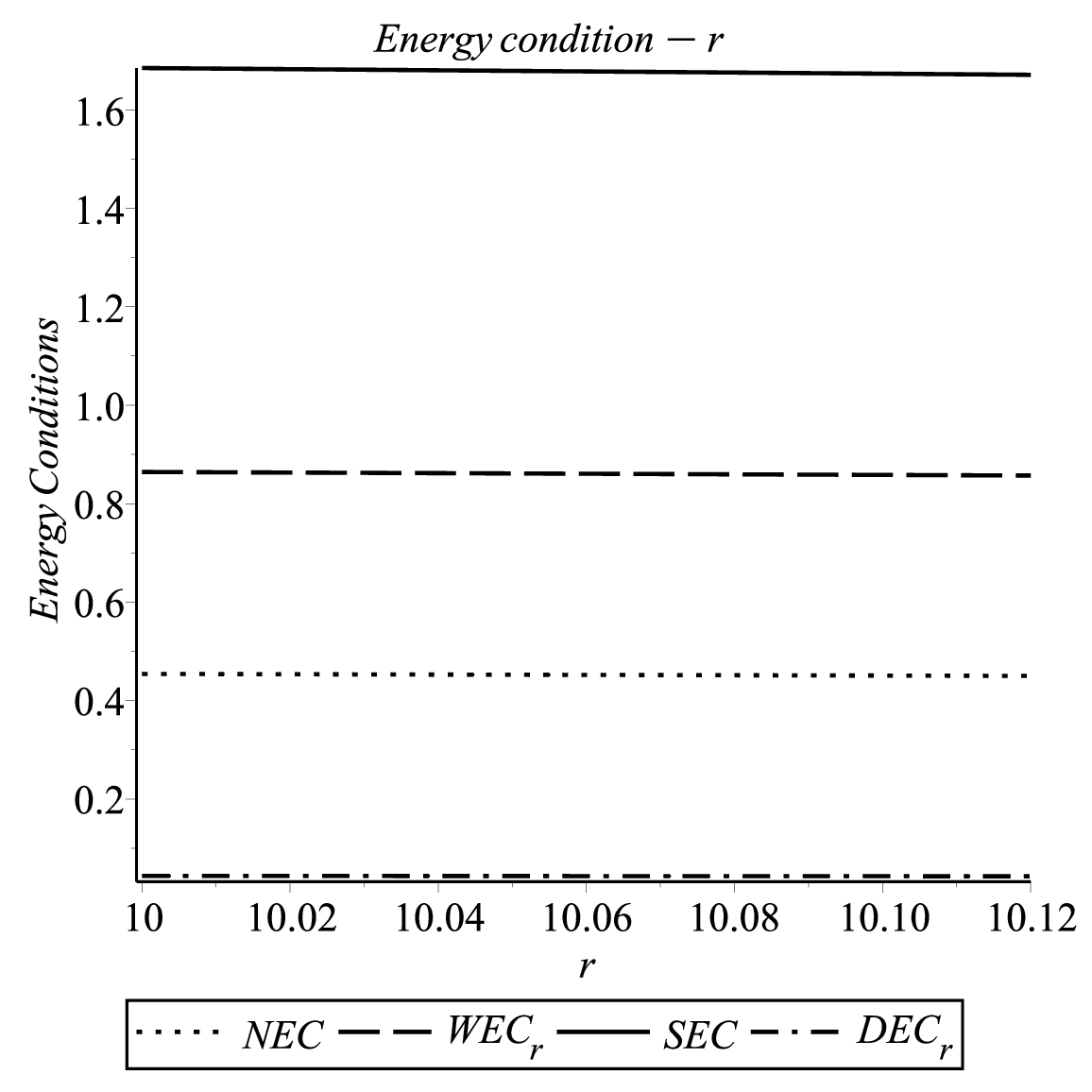}
\caption{Variation of the different energy conditions w.r.t. the radial coordinate $r$.}\label{pres.}
\end{figure}
%%%%%%%%%%%%%%%%%%%%%%%%%%%%%%%%%%%%%%%%%%%%%%%%%

\subsection{Herrera's cracking condition}

Cracking or breaking is a phenomenon in compact objects resulting as a consequence of development of perturbations in a spherically symmetric matter distribution at equilibrium, due to assymetry in radial forces in different regions. The most common cause of this is thought to be the presence of anisotropy in the constituting fluid. The sound speed squared in\cite{Herrera1992} both the tangential and radial directions must be a positive number less than unity in order to prevent violation of causality and additionally according to Herrera~\cite{Herrera1992} and Andreasson~\cite{Andreasson2009}, the absolute value of the difference in the squares of the radial and tangential sound speeds must also be less than or equal to unity to prevent cracking which makes the system unstable. However, as already discussed, the modified field equations in our model present themselves with an inherent isotropy due to equality of the effective pressure terms in the radial and tangential directions.  So, it is unusual that our gravastar model develops cracking due to effective fluid anisotropy. However, the sound speed squared must be less than unity throughout the shell to prevent development of perturbations leading to cracking.

%%%%%%%%%%%%%%%%%%%%%%%%%%%%%%%%%%%%%%%%%%%%%%%%%%%%%%%%%%%%
\begin{figure}[!htp]
	\centering
	\includegraphics[width=5cm]{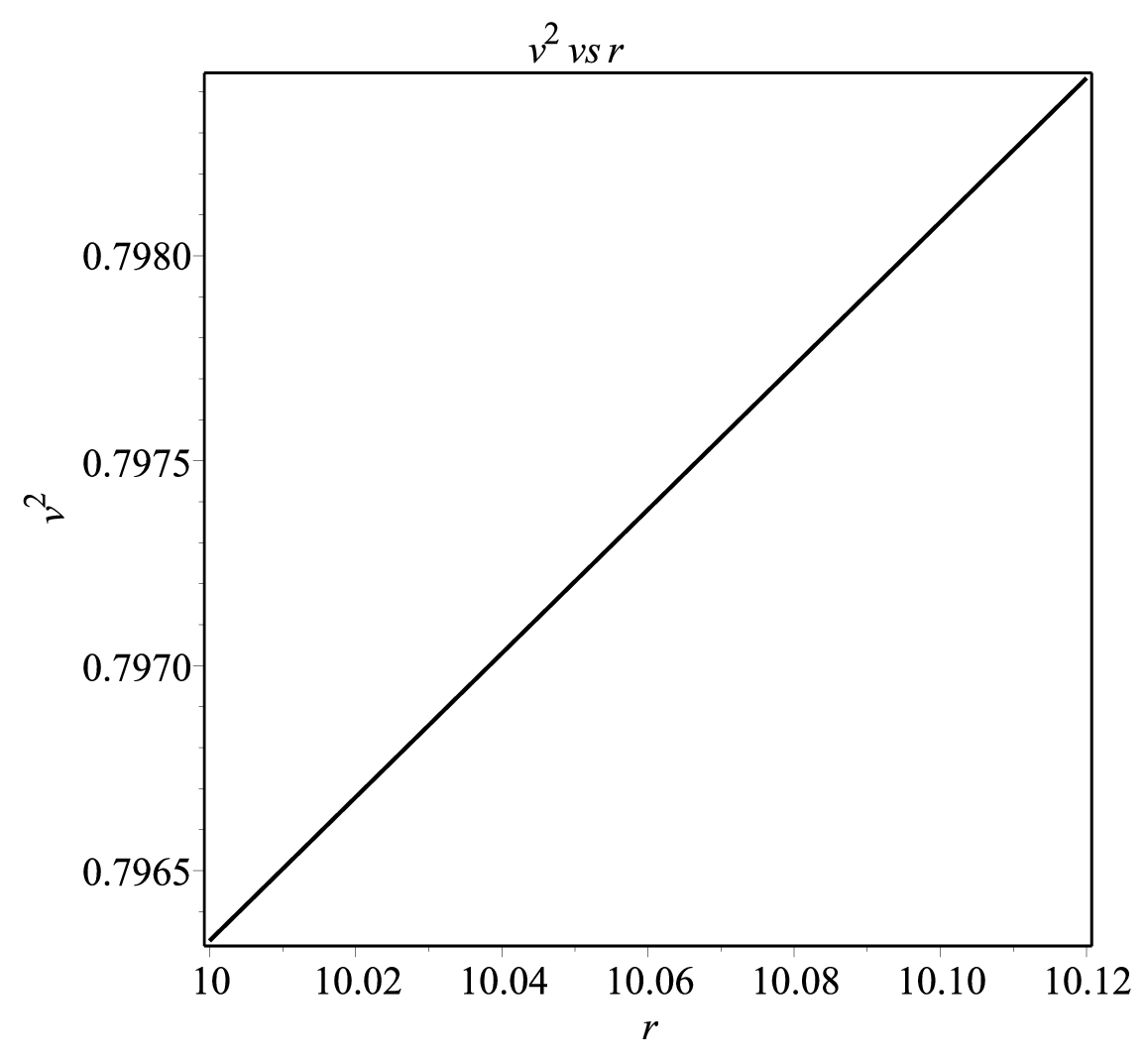}
	\caption{Variation of $v^2_{s}$ w.r.t. the radial coordinate $r$.}\label{vel.}
\end{figure}
%%%%%%%%%%%%%%%%%%%%%%%%%%%%%%%%%%%%%%%%%%%%%%%%%%%%%%%%%%%

We obtain the sound speed squared in the gravastar shell as    
\begin{equation}
v^2_{s}=\frac{dp^{eff}}{d\rho^{eff}}=\frac{6p_0e^{-Br^2}-\rho_c}{2p_0e^{-Br^2}-\rho_c}  \label{eq40} \\
\end{equation}

The variation of the square of sound speed along the radial expanse of the shell has been plotted in Figure 10. The sound speed is minimum at the interior boundary and maximum at the exterior boundary of the gravastar, increasing montonically along the shell. We find it to be well behaved and less than unity across the entire shell width. Thus, our gravastar model passes all three tests for stability and can be considered to be stable.  

\section{Discussion and Conclusion}

We have constructed a stable gravastar model in the framework of LQC in four spacetime dimensions. The background theory is known to be a well accepted correction to standard GR at the UV scale characterized by high energy densities which can be naturally expected in compact objects like gravastars. It is due to the prevelant high energy densities that the quadratic corrected $\rho^2$ term in the modified EFE becomes significant at the UV scale and can later be ignored at lower energy densities, leading to the recovery of standard GR.  There is an inherent pressure isotropy present in the modified EFE. As shown by Cattoen~\cite{Cattoen}, stable gravastars must have a pressure anisotropy in order to be stable in the standard relativistic context, but we successfully obtain a stable gravastar model in the absence of any pressure anisotropy due to the presence of the UV corrected term in the modified EFE from the background theoretical framework. This is one of the novel features of our gravastar model. 

We move on to obtaining the unknown radial metric potential for the interior and the shell of the gravastar. The solutions are well behaved throughout the corresponding regions of the gravastar and there is no central singularity present. The matter comprising the gravastar interior is a gravitational BEC resulting from a gravitational quantum phase transition with the shell acting as a critical surface, which is characterized by an EoS $p=-\rho$ indicating gravitationally repulsive matter violating the SEC. However, this does not gurantee the violation of NEC at the interior. The central singularity in a self-gravitating object of extremely high energy density is basically a curvature singularity of the Ricci type, which is an essential physical singularity that cannot be removed by a co-ordinate transformation, as such a singularity is characterized by a diverging Riemann curvature tensor resulting from a diverging energy density. 

The most unique feature of our gravastar model that we shall justify here is as follows. The modified field equations that we consider in the frameworks of LQC has a very interesting feature. There is a maximum critical density to which the energy density of the space-time can rise. This density $\rho_c$, beyond which the density cannot rise further and the singularity is replaced by a bounce. This can be verified from the value we have obtained for the constant interior density of the gravastar. This value turns out to be identical to the value of the critical density $\rho_c=0.41m^4$. Here the value is in units of Planck mass and in LQC framework, the value is higher as the Planck mass is in four dimensions and hence in GeV scale. As the density of the gravastar interior equals that of the critical density, the density cannot exceed further and hence cannot diverge, resulting in a non-diverging (finite) Riemann curvature tensor and a corresponding non-singular spacetime, such that the central singularity of the self gravitating object is replaced by a non-singular (regular) bounce that facilitates the formation of the gravastar in the underlying UV corrected gravity framework. The incapability of the energy density to rise above the critical value may be physically attributed to a gravitationally repulsive force, that is both quantitatively and qualitatively different from dark energy which is basically a correction to standard GR at the infra red (IR) scale. It also does not involve any electromagnetic source. This \textit{repulsive force} can be thought to arise as a consequence of \textit{the quantization of space-time itself in the LQC context.}       

Another key feature of our gravastar model is that no thin shell approximation is required to be enforced to solve the modified EFE analytically in the shell region. This results in the replacement of the infinitesimal thin shell by a more realistic and physically justifiable thin shell of finite thickness. Various physical properties of the shell have been studied including the energy density and pressure of the shell, the total energy of the shell, the entropy of the shell and its proper thickness. The variation of these physical properties across the radial expanse of the shell have been found to be continuous and well behaved. The surface density and surface pressure of the gravastar have been obtained from a physical point of view, as the presence of matter on the gravastar surface that can be inferred from a non zero energy density at the exterior surface of the gravastar results in an extrinsic discontinuity leading to a finite surface density and pressure. These quantities have been computed making use of the well known Israel-Darmois junction conditions and vanishing of these quantities at the exterior boundary of the gravastar along with the matching conditions obtained from the analytic continuation of the metric potentials and their first derivatives across the interior and exterior surface boundaries, provide us with the equations for obtaining the various unknown constant model parameters which are extremely important for our analysis, as we have shown in the paper. Cursory stability checks have been performed successfully for our gravastar model using well accepted approaches like computing the surface redshift, checking the validity of the energy conditions for the effective matter description and checking that causality holds good via the validity of the reduced Herrera cracking condition for the inherently isotropic effective matter distribution.

We discuss very briefly a few ideas on the possibility of detection of gravastars. Specifically speaking in the context of the braneworld scenario, we note that the braneworld embedded in a higher dimensional bulk geometry has been found relevant in the context of explaining the recent detections of the Gravitational wave (GW) event GW170817 associated with the LIGO/Virgo collaboration and its electromagnetic counterpart in the form of gamma-ray burst GRB170817A associated with the Fermi Gamma-ray Burst Monitor and the INTEGRAL Anti-Coincidence Shield spectrometer~\cite{Visinelli}. Also, the dark shadow of M87$^{\ast}$ can be explained using a braneworld scenario with an Anti-deSitter($AdS_5$) bulk geometry~\cite{Vagnozzi}. This motivates us to study gravastar in a framework that is an accepted UV corrected theory of GR and has found relevance in describing non-singular black hole bounce\cite{lqc} and non-singular bouncing models of the universe which are cyclic in nature\cite{LQC}. In general, methods of possible indirect detection of gravastar have been discussed in literature and involve the study of gravastar shadows~\cite{Sakai2014}, gravastar microlensing effects of higher maximal luminosity as compared to equivalent mass black holes\cite{Kubo2016}, possible generation of  ringdown signal of $GW~150914$~\cite{Abbott2016} by a horizonless object like gravastar~\cite{Cardoso1,Cardoso2}. We conclude the paper with the notion that \textit{LQC appears to be an idealized framework for describing non-singular self gravitating end state objects like gravastars.}

\section*{Acknowledgement}
The authors gratefully acknowledge support from the Inter-University Centre for Astronomy and Astrophysics (IUCAA), Pune, India for warm hospitality and allowing to use the research facilities, where this work was done.

\end{document}